\documentclass[aps,prl,reprint]{revtex4-1}
\usepackage{graphicx}

\begin{document}
\title{Dynamics of stress $p53$: Nitric oxide induced transition of states and synchronization}
\author{Gurumayum Reenaroy Devi$^1$}
\author{Md. Jahoor Alam$^1$}
\author{R. Ramaswamy$^2$}
\author{R.K. Brojen Singh$^{3}$}
\email{brojen@jnu.ac.in}
\affiliation{$^1$Centre for Interdisciplinary Research in Basic Sciences, Jamia Millia Islamia,
New Delhi 110025, India\\$^2$School of Physical science, Jawaharlal Nehru University, New Delhi 110067, India\\$^3$School of Computational and Integrative Sciences, Jawaharlal Nehru University, New Delhi-110067, India}

\begin{abstract}
We study the temporal and the synchronous behaviours in $p53-Mdm2$ regulatory network due to the interaction of its complex network components with the nitric oxide molecule. In single cell process, increase in nitric oxide concentration gives rise the transition to various $p53$ temporal behaviours, namely fixed point oscillation, damped oscillation and sustain oscillation indicating stability, weakly activated and strongly activated states. The noise in stochastic system is found to help to reach these states much faster as compared to deterministic case which is evident from permutation entropy dynamics. In coupled system with nitric oxide as diffusively coupling molecule, we found nitric oxide as strong coupling molecule within a certain range of coupling strength $\epsilon$ beyond which it become weak synchronizing agent. We study these effects by using correlation like synchronization indicator $\gamma$ obtained from permutation entropies of the coupled system, and found five important regimes in ($\epsilon-\gamma$) phase diagram, indicating desynchronized, transition, strongly synchronized, moderately synchronized and weakly synchronized regimes respectively. We claim that there is the competition between the toxicity and the synchronizing role of nitric oxide that lead the cell in different stressed conditions. 
\end{abstract}

\maketitle

\section{Introduction}
$p53$ is an intrinsic protein in the biological cells. It is associated with more than 50 percent of the human cancers. It is involved in many key  metabolic pathway regulations such as tumor suppression, cell cycle arrest, DNA repair and apoptosis \cite {lane,chi}. $p53$ protein level  is believed to be always fluctuating within the cell because of its participation in various networks. Several studies have been performed so far for the understanding of the fluctuation of $p53$ protein within the biological cell which reveals that it is the main controller of the cellular functions. One of the key protein which directly associated with the dynamics of $p53$ is $Mdm2$ protein. $Mdm2$ is a negative feedback regulator of $p53$ protein \cite{lane,Geva}. In an unstressed cell $Mdm2$ controls the level of $p53$ \cite {mom}. However, different models have also been developed on the dynamics of $p53-Mdm2$ pathway but it remains unclear about many unknown factors which are still responsible for changes in the dynamics of $p53-Mdm2$ pathway.

Nitric oxide ($NO$) is an important, extremely short lived and bioactive molecule ($\sim 1-10~seconds$) \cite{sch,stern} which can trigger various physiological and pathological processes in a wide variety of mammalian cell types \cite{xin}. It is widely and actively synthesized by various $NO$ synthase enzymes (NOS), namely neuronal (nNOS), inducible (iNOS) \cite{low} or endothelial (eNOS) \cite{li,wer} such that these isoforms convert arginine to $NO$ and citruline \cite{low,li,wer,mar,din}. It has two contrast roles in different single cell types, the first one is it induces apoptosis (programmed cell death) in some cell types such as macrophages, neurons, pancreatic $\beta$-cells, thymocytes, chondrocytes, hepatocytes \cite{chu,bru,kim} etc, whereas the second one is it inhibits apoptosis in other cell types such as B-lymphocytes, eosenophils, ovarian follicles, neuronal PC12 cells, embryonic motor neurons \cite{li1, wan,tay,kim1} etc. Further, it is reported that $NO$ induced apoptotic signaling pathways in human lymphoblastoid cell harboring $p53$ protein \cite{oka}.

Other important functions of $NO$ are its ability to induce cellular stress, activation of $p53$ via DNA damage and disruption of energy metabolism, calcium homeostasis and mitochondrial function which can be taken as toxic action that leads to cell death \cite{hof,hus,chu1,ker,hal,lei,mur}. It is achieved by upregulating $p53$ \cite{mes,bru} and downregulating $Mdm2$ \cite{wan} via DNA damage induced by $NO$ causing growth arrest in cell cycle by giving time for DNA repair \cite{lev}. This means that increase in nitric oxide in a cell also induce increase in toxic in the cell. Several experimental studies shows that nitric oxide acts as a regulatory factor for $Mdm2$ protein which ultimately leads to the fluctuation of the $p53$ \cite{hof,mes,wan}. Extremely excess of $NO$ may lead $p53$ to cause cell apoptosis \cite{bru}. 

One of the most important role of $NO$ is its ability to act as an excellent intercellular signaling molecule \cite{mar,din}. The reason could be $NO$ is small and hydrophobic molecule which can pass through cell membrane easily and it is actively and abundantly created inside the cell \cite{din,mur} and again it can also diffuse through several cell diameters from its site of synthesis \cite{mur,lan1,lan2}. This diffusion of $NO$ can lead to various intracellular signal processing and intercellular communication. Further, this diffusion and intracellular consumption are the two main factors which control $NO$ concentration level in biological cells \cite{ded,che}.

There are various issues which are still not fully resolved. How $NO$ level is maintained inside the cell since it is toxic in some cell types, whereas this level can prevent apoptosis to some others, is not fully resolved. Even if $NO$ induce toxic to cells, how does it activate $p53$ leading to cellular stress and excess stress cause apoptosis, is still need to be investigated. Further, even if $NO$ is considered as synchronizing molecule, what could be its role in coupling $p53-Mdm2$ oscillators at different stress conditions, is still need to be investigated and resolved. In this work we study an integrated model of intracellular $p53-Mdm2$ oscillator with $NO$ synthesis pathway to resolve some of the issues mentioned. The stressed $p53-Mdm2$ oscillators induced by $NO$ are diffusively coupled via $NO$ and investigated the impact of $NO$ on single $p53-Mdm2$ network dynamics and on the rate of synchronization among the coupled oscillators.
\begin{table*}
\begin{center}
\begin{tabular}{c c c c}
{\bf Table 1 - List of molecular species}\\ \hline
{\bf S.No}	&        {\bf  Species Name} 	&     {\bf Description}         &  {\bf Notation} 	\\ \hline
1.		& 		p53 		& 	Unbound p53 protein 	& 	$x_1$ 		\\ 
2. 		& 		Mdm2 		& 	Unbound Mdm2 protein 	& 	$x_2$ 		\\ 
3. 		& 		$Mdm2\_p53$ 	& 	Mdm2/p53 complex 	& 	$x_3$ 		\\ 
4. 		& 		$Mdm2\_mRNA$ 	& 	Mdm2 messenger RNA 	& 	$x_4$ 		\\
5. 		& 		$NO$ 		& 	Unbound Nitric Oxide  	& 	$x_5$ 		\\ 
6. 		& 		$NO\_Mdm2$ 	& 	NO/Mdm2 complex 	& 	$x_6$ 		\\ \hline       
\end{tabular}
\end{center}
\end{table*}

\begin{table*}
\begin{center}
{\bf Table 2 List of Chemical Reactions, Propensity Function(P.F.) and Rate constants}
 \begin{tabular}{c rrrrr}
\\ \hline

        {\bf Sl.No} & {\bf Reaction } &  {\bf P.F.} & {\bf Rate Constants}&{\bf Reference}\\ \hline
        1  & $x_4\stackrel{k_1}{\longrightarrow}x_4+x_2$ &  $k_1 x_4 $ & $4.95\times10^{-4} sec^{-1}$ & \cite{pro,fin} \\ 
        2  & $x_1\stackrel{k_2}{\longrightarrow}x_1+x_4$ &  $k_2 x_1 $ & $1.0\times 10^{-4} sec^{-1}$ & \cite{pro,fin} \\ 
        3  & $x_4\stackrel{k_3}{\longrightarrow}\phi$ & $k_3 x_4 $ & $1.0\times 10^{-4} sec^{-1}$ & \cite{pro,fin}\\ 
        4  & $x_2\stackrel{k_4}{\longrightarrow}\phi$ & $k_4 x_2 $ & $4.33\times 10^{-4} sec{-1}$ & \cite{pro,fin}  \\
        5  & $\phi\stackrel{k_5}{\longrightarrow}x_1$ &  $k_5$ & $0.78 sec^{-1}$ & \cite{pro} \\ 
        6  & $x_3\stackrel{k_6}{\longrightarrow}x_2$ &  $ k_6 x_3 $ & $8.25\times 10^{-4} sec^{-1}$ & \cite{pro}  \\ 
        7  & $x_1+x_2\stackrel{k_7}{\longrightarrow}x_3$ &  $k_7 x_1 x_2 $ & $11.55\times 10^{-4} mol^{-1} sec^{-1}$ & \cite{pro}  \\ 
        8  & $x_3\stackrel{k_8}{\longrightarrow}x_1+x_2$ &  $k_8 x_3 $ &$11.55\times 10^{-6} sec^{-1}$ & \cite{pro,fin}  \\       
        9  & $\phi\stackrel{k_{9}}{\longrightarrow}x_5$ & $k_{9}$ & $1\times 10^{-2} mol^{-1} sec^{-1}$ & \cite{wang,jah}  \\ 
        10 & $x_5+x_2\stackrel{k_{10}}{\longrightarrow}x_6$ & $k_{10} x_5 x_2$ & $1\times 10^{-3} mol^{-1} sec^{-1}$ & \cite{jah} \\ 
        11 & $x_6\stackrel{k_{11}}{\longrightarrow}x_5$ & $k_{11} x_6$ & $3.3\times 10^{-4} sec^{-1}$ & \cite{wang,jah}\\ 
        12 & $x_5\stackrel{k_{12}}{\longrightarrow}\phi$ & $k_{12} x_5$ & $1 \times 10^{-3} sec^{-1}$ & \cite{wang,jah}\\ \hline
        
\end{tabular}
\end{center}
\end{table*} 

\section{Materials and Methods}

\subsection{The stress p53-Mdm2 oscillator induced by NO} 

Nitric oxide ($NO$) can diffuse across the cell membrane \cite{stern} and is constantly produced in the cell through enzyme metabolism \cite{stern,wood}. Recent studies shows that nitric oxide down regulates the $Mdm2$ protein \cite{wang,sci}. Down regulation of $Mdm2$ protein leads  to the fluctuation of $p53$ protein \cite {stern}. We consider $Mdm2$ as well as $p53$ proteins moves in and out of the nucleus. These proteins after activation localized in the nucleus and activate target genes \cite{chen,lia}. $p53$ transcriptionally activates $Mdm2$ gene to form $Mdm2\_mRNA$  due to which production of $Mdm2$ protein increases in the cells. $Mdm2$ forms complex with $p53$ \cite{pro}. After forming the complex $Mdm2$ ubiquitinates $p53$ due to which the $p53$ is degraded\cite{haup,kubb,moma}. $NO$ forms the complex with cytosolic $Mdm2$ protein due to which the $Mdm2$ protein is degraded\cite {wang,sci}. With downregulation of $Mdm2$ protein, $p53$ is also fluctuated and it shows oscillatory behavior\cite{jah}. The life time of $p53$ is very short with half life of around 30 minutes \cite{fin}. Further the life span of $Mdm2$ protein, $Mdm2\_mRNA$ and $NO$ are very short with half life periods around 30 minutes \cite{fin,pan}, $60-120$ minutes \cite{hsi,men} and 5-10 seconds \cite{wood,wang} respectively. These molecules are regulated inside the cell itself from time to time. Consequently $p53$ is an integral protein in the cell and genetically regulated constantly inside the cell \cite {bride} keeping its population stabilized at low level in normal cells and it is also connected with huge number of sub-cellular networks. The biochemical reaction network model is shown in Fig. 1.  Here we have symbolized the molecular species in terms of x 's for the sake of simplicity in the calculation and their symbols are shown in Table 1. The corresponding reaction channels with their respective transition rates are shown in Table 2.
\begin{figure}
\label{}
\begin{center}
\includegraphics[height=270 pt,width=8cm]{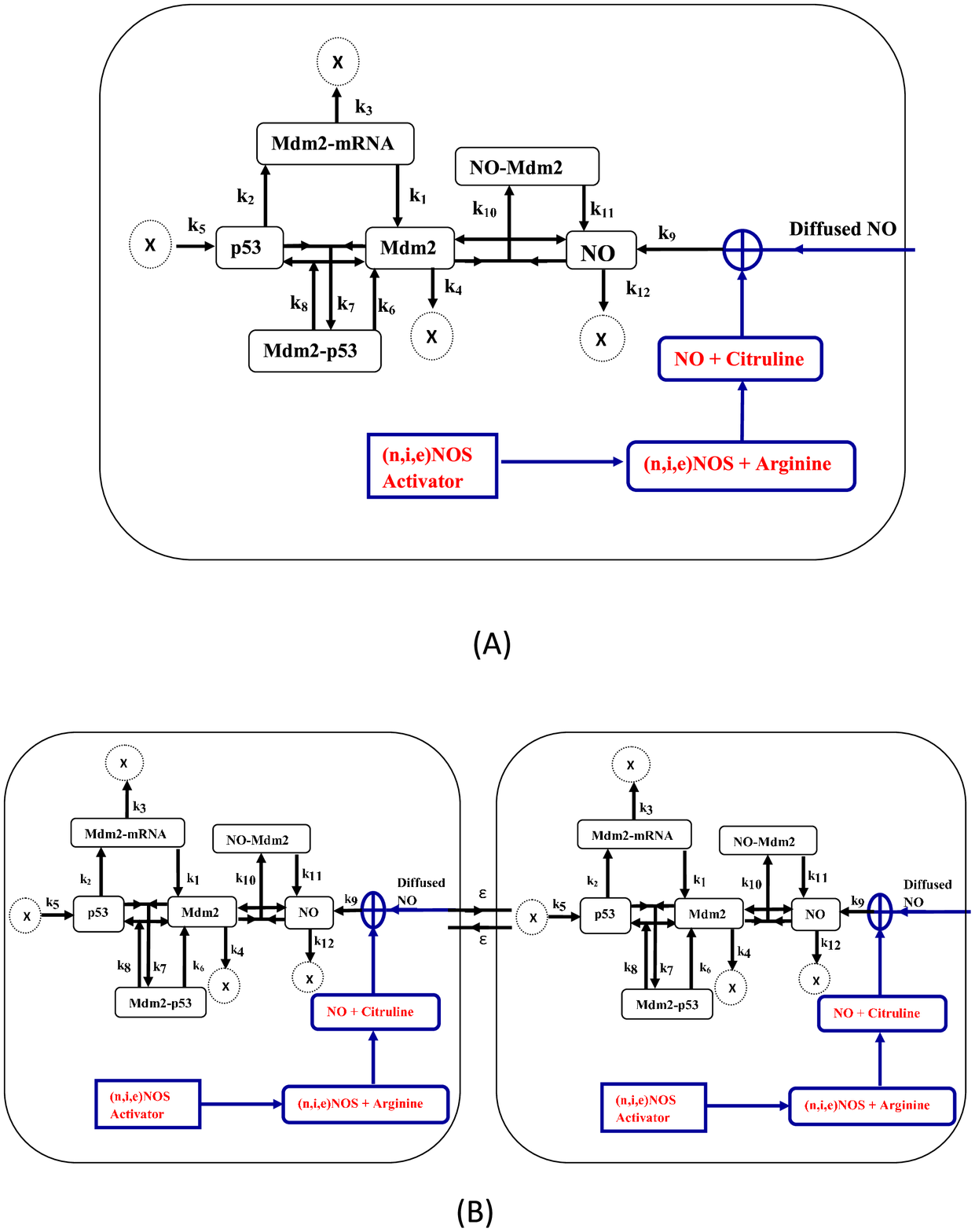}
\caption{[A]A schematic diagram of reaction network model of p53-Mdm2 Oscillator induced by NO.[B]A schematic diagram of two identical oscillators diffusively coupled with NO.} 
\end{center}
\end{figure}

In deterministic system, the biochemical reactions shown in Fig. 1 can be translated into a set of coupled ordinary differential equations using simple Mass-action kinetic law. We denote $p53$ as $x_1$, $Mdm2$ as $x_2$, $Mdm2\_p53$ complex as $x_3$, $Mdm2\_mRNA$ as $x_4$, $NO$ (nitric oxide) as $x_5$, and $NO\_Mdm2$ complex as $x_6$. The equations are given by,
\begin{eqnarray}
\frac{dx_1}{dt}&=& k_5-k_7x_1x_2+k_8x_3\\
\frac{dx_2}{dt}&=& k_1x_4-k_4x_2+k_6x_3-k_7x_1x_2+k_8x_3\nonumber\\
&&-k_{10}x_5x_2\\
\frac{dx_3}{dt}&=&-k_6x_3+k_7x_1x_2-k_8x_3\\
\frac{dx_4}{dt}&=& k_2x_1-k_3x_4\\
\frac{dx_5}{dt}&=& k_9-k_{10}x_5x_2+k_{11}x_6-k_{12}x_5\\
\frac{dx_6}{dt}&=& k_{10}x_5x_2-k_{11}x_6
\end{eqnarray}
Cellular and sub-cellular processes are complex stochastic or noise induced processes due to random molecular interaction in the system \cite{rao} and system interaction with the environment \cite{mca,bla}. Stochastic model which is a realistic model with qualitative and quantitative perscriptions, can be well described by taking each and every molecular interaction systematically to find their trajectories in configuration space \cite{gill}. This can be done by constructing Master equation of the interaction network, which is mathematically the time evolution of configurational probability $P(\vec{x},t)$ with $\vec{x}=(x_1,x_2,\dots,x_6)^{-1}$ based on decay and creation of each molecular species at each molecular interaction \cite{gill,mcq}. However, it is very difficult to solve Master equation for complex systems except for simple ones. Computationally one can compute the trajectory of each and every molecular species in the system using stochastic simulation algorithm (SSA) due to Gillespie \cite{gill} by taking every possible interaction in the complete system. Further, one can simplify this Master equation based on some realistic assumptions which are small time interval of any two consecutive interactions and large molecular population limit \cite{gill1}. This let the Master equation to reduce to simpler Chemical Langevin equations (CLE). For our system, we have following CLEs,

\begin{eqnarray}
\frac{dx_1}{dt}&=& k_5-k_7x_1x_2+k_8x_3\nonumber\\
&&+\frac{1}{\sqrt{V}}\left[\sqrt{k_5}\xi_1-\sqrt{k_7x_1x_2}\xi_2+\sqrt{k_8x_3}\xi_3\right]\\
\frac{dx_2}{dt}&=& k_1x_4-k_4x_2+k_6x_3-k_7x_1x_2+k_8x_3\nonumber\\
&&-k_{10}x_5x_2+\frac{1}{\sqrt{V}}\left[ \sqrt{k_1x_4}\xi_4-\sqrt{k_4x_2}\xi_5\right]\nonumber\\
&&+\frac{1}{\sqrt{V}}\left[\sqrt{k_6x_3}\xi_6-\sqrt{k_7x_1x_2}\xi_7+\sqrt{k_8x_3}\xi_8\right]\nonumber\\
&&-\frac{1}{\sqrt{V}}\left[\sqrt{k_{10}x_5x_2}\xi_9\right]\\
\frac{dx_3}{dt}&=&-k_6x_3+k_7x_1x_2-k_8x_3-\frac{1}{\sqrt{V}}\left[\sqrt{k_6x_3}\xi_{10}\right]\nonumber\\
&&+\frac{1}{\sqrt{V}}\left[\sqrt{k_7x_1x_2}\xi_{11}-\sqrt{k_8x_3}\xi_{12}\right]\\
\frac{dx_4}{dt}&=& k_2x_1-k_3x_4\nonumber\\
&&+\frac{1}{\sqrt{V}}\left[\sqrt{k_2x_1}\xi_{13}-\sqrt{k_3x_4}\xi_{14}\right]\\
\frac{dx_5}{dt}&=& k_9-k_{10}x_5x_2+k_{11}x_6-k_{12}x_5\nonumber\\
&&+\frac{1}{\sqrt{V}}\left[\sqrt{k_9}\xi_{15}-\sqrt{k_{10}x_5x_2}\xi_{16}\right]\nonumber\\
&&+\frac{1}{\sqrt{V}}\left[\sqrt{k_{11}x_6}\xi_{17}-\sqrt{k_{12}x_5}\xi_{18}\right]\\
\frac{dx_6}{dt}&=& k_{10}x_5x_2-k_{11}x_6\nonumber\\
&&+\frac{1}{\sqrt{V}}\left[\sqrt{k_{10}x_5x_2}\xi_{19}-\sqrt{k_{11}x_6}\xi_{20}\right]
\end{eqnarray}
where, $V$ is the system size and $\xi_i$, $i=1,2,\dots,20$ are random noise parameters which are given by, $\xi_i(t)\xi_j(t^\prime)=\delta_{ij}\delta(t-t^\prime)$. The noise term varies with order $O(V^{-1/2})$.
\begin{figure}
\label{}
\begin{center}
\includegraphics[height=230 pt,width=7.5cm]{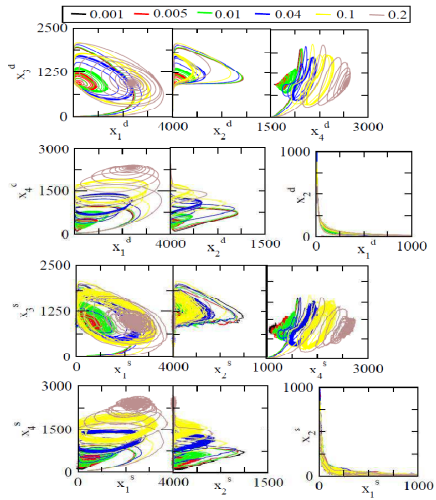}
\caption{The 2D plots for different proteins and their complexes for different values of $k_{NO}$ 0.001,0.005,0.01,0.04,0.1,0.2.} 
\end{center}
\end{figure}

\subsection{Numerical techniques}
The deterministic set of differential equations: $\frac{dx_i}{dt}=G_i(x_1,x_2,...,x_M);~i=1,2,...,M$, and set of Chemical Langevin equations: $\frac{dx_i}{dt}=G_i(x_1,x_2,...,x_M)+\beta_i(\xi_1,\xi_2,...)U_i(x_1,x_2,...,x_M);~i=1,2,...,M$ can be solved using standard 4th order Runge-Kutta algorithm for numerical integration \cite{press}. Here, $G_i$, $\beta_i$ and $U_i$ are some functions. The parameters needed in the differential equations are obtained from various experimental works reported which are listed in Table 2. Uniform random number generator which generate random numbers between 0 and 1 is used in the case of solving CLE. We wrote our own code in java for simulation purpose \cite{ Her}.

We use stochastic simulation algorithm (SSA) due to Gillespie \cite{gill} to simulate the biochemical reaction network thereby to understand the dynamical behaviors of each participating molecular species in the system. The algorithm is a Monte Carlo type and is based on the basic fact that the trajectory of each species can be traced out if one understands which reaction is fired at what time. The technique uses two uniform random number generators, one for identifying reaction number fired and the other to pick up time of reaction fired.

\subsection{Measuring complexity: Permutation entropy}
To understand the complexity and information contain in the dynamics of the $p53$ and $Mdm2$, we calculate permutation entropy $H$ of each variable dynamics \cite{ban} for the various $k_{NO}$ values taken both in deterministic and stochastic systems. The permutation entropy spectrum of a variable $x(t)$ can be calculated by mapping it onto a symbolic sequence of length $N$: $x(t)=\{x_1,x_2,\dots,x_N\}$ \cite{ban,cao}. The sequence is then partitioned into $M$ number of short sequences of equal size $L$ each such that, $x(t)=\{q_1,q_2,\dots,q_M\}$ with $q_i=\{x_{i+1},x_{i+2},\dots,x_{i+L}\}$ and by sliding this window of size $L$ with maximum overlapping. The permutation entropy of any short sequence $q_i$ can be calculated by defining a r-dimensional space, $U_i=\{x_{i+1},x_{i+2},\dots,x_{i+r}\}$ with embedded dimension $r$, finding out all possible inequalities of dimension $r$ and mapping the inequalities along $q_i$ in ascending order to obtain probabilities of occurrence of each inequalities ($p_j:j=1,2,...$). Since only $S$ out of $r!$ permutations are distinct one can define normalized permutation entropy by, $H_i(r)= - \frac{1}{ln(r!)}\sum_{j=1}^{S}p_jlnp_j$ where, $0\le H_i(r)\le 1$, and permutation entropy spectrum of the variable $x(t)$ is given by $H(t)=\{H_i:i=1,2,\dots,M\}$. This $H(t)$ will measure the complexity of the data $x(t)$.
\begin{figure}
\label{}
\begin{center}
\includegraphics[height=260 pt,width=7.5cm]{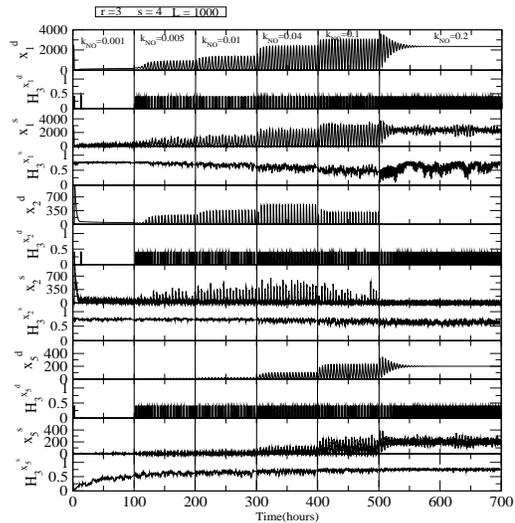}
\caption{Plots of p53 and Mdm2 activation via various concentration levels NO (indicated by NO creation rate constant) and their respective permutation entropies.} 
\end{center}
\end{figure}

In general noise enhances $H(t)$ that leads to increase in complexity in the dynamics, however there are cases where noise reduces $H(t)$ value \cite{ban}. But if the strength of the noise is small enough, it does not cause significant change in complexity in the dynamics \cite{ban}. The stochastic dynamics are noise induced dynamics \cite{mca, bla, mcq, gill, ram, kam} where the strength of the noise depends on system size etc. Further noise has two distinct contrast roles in dynamical systems, if the strength of the noise is compartively small (smaller than some defined critical value $\Gamma^C$ that may be different for different systems) then it induces order (decreasing complexity) to carry out important constructive functions known as stochastic resonance \cite{ben,wie,gam,ani}, and if the strength of noise is comparatively large (larger than $\Gamma^C$) it hinderances the dynamics enhancing disorderness (increasing complexity). This gives us a notion that noise has an important impact on $H(t)$ in stochastic dynamics.

In stochastic system each element in symbolic sequence $x(t)$ can be expressed as $x_i^s=x_i\pm\Gamma_i\sigma_i$, where, $i=1,2,...,N$, $\sigma_i$ is random parameter with $\langle\sigma_i\rangle=0, $ but $\langle\sigma_i\sigma_j\rangle=1$ for $i=j$ but 0 for $i\ne j$, $\Gamma_i$ is noise strength and superscript $s$ indicates stochastic element \cite{kam}. For $r=2$, there are two distinct possible inequalities or states $x_i^s\langle x_{i+1}^s$ and $x_i^s\rangle x_{i+1}^s$. If we take $\Delta x^s=x_i^s-x_{i+1}^s$ and $\Delta x=x_i-x_{i+1}$ then we have $\frac{\Delta^2x}{\Gamma}=\frac{\Delta x^s-\Delta x}{\Gamma}\approx\pm\Delta\sigma$, where $\Delta\sigma=\sigma_i-\sigma_{j+1}$ and $\Gamma_i\approx\Gamma_{i+1}\approx\Gamma$ is taken. Since switching to any one of the two states depends on $\Delta\sigma$ (depending on the sign) and $\Delta\sigma$ is random in nature in the time series, $\Delta\sigma$ could be think of as a random switching parameter. Therefore, $H^s(t)=\{H^s_j:j=1,2,\dots,M\}$ is a stochastic spectrum and is a noise induced process. For small noise strength ($\frac{\Delta^2x}{\Gamma}\rangle\rangle \Delta\sigma$) this random switching mechanism may not active, and therefore this stochastic spectrum may recover classical behaviour $\langle\frac{\Delta^2x}{\Gamma}\rangle\rightarrow 0$ such that $\langle\Delta x^s\rangle\rightarrow\langle x\rangle=x_A$, such that $H^s(t)\rightarrow H(t)$. Hence, a small noise does not give much impact on $H^s(t)$ spectrum.

However, if $\frac{\Delta x}{\Gamma}$ is comparable to $\Delta\sigma$, $H^s(t)$ is very much affected by noise because there is competition between $\frac{\Delta x}{\Gamma}$ and $\Delta\sigma$ such that switching mechanism from one distinct state to another becomes active that leads to $H^s(t)$ a different spectrum. Therefore, at this condition $H^s(t)$ may be quite different from $H(t)$, and so it could give $\langle\frac{\Delta^2x}{\Gamma}\rangle_E\ne 0$ but the ensemble average (denoted by subscript $E$) will reduce the fluctuation but not the dynamics.
\begin{figure}
\label{}
\begin{center}
\includegraphics[height=270 pt,width=7.5cm]{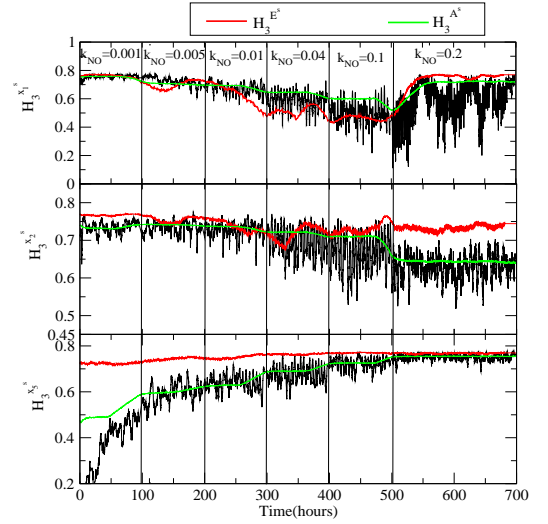}
\caption{Comparision of permutation entropy spectrums of (A) single time series data, (B) averaging of time series data then calculate permutation entropy, and (C) calculate permutation entropy spectrums of each time series data and then average.} 
\end{center}
\end{figure}

\subsection{Detection of synchronization}
The measure of synchrony for the two coupled systems can be done by the permutation entropy method \cite{liu}. The method allows to define the permutation entropies of $x_1^{[1]}$ and $x_1^{[2]}$ to be $H_{x_1^{[1]}}(r)=-\frac{1}{ln(r!)}\sum_{j=1}^{L}p_jlnp_j$ and $H_{x_1^{[2]}}(r)=-\frac{1}{ln(r!)}\sum_{j=1}^{L}p_jlnp_j$ respectively. This leads us to write back the variables as $x_1^{[1]}(t)=\{H_{x_1^{[1]}}^{[1]},H_{x_1^{[1]}}^{[2]},\dots,H_{x_1^{[1]}}^{[M]}\}$ and $x_1^{[2]}(t)=\{H_{x_1^{[2]}}^{[1]},H_{x_1^{[2]}}^{[2]},\dots,H_{x_1^{[2]}}^{[M]}\}$ respectively. Then a correlation like function $C(r_j)$ can be defined as,
\begin{eqnarray}
C(r_j)&=&1~~~~H_j(r)\rangle H_{j-1}(r)\nonumber\\
&=&-1~~~~otherwise
\end{eqnarray}
Now for the two systems $C_{x_1^{[1]}}(r)$ and $C_{x_1^{[2]}}(r)$ are calculated in the same manner to define an order parameter $\gamma$ to measure rate of synchronization,
\begin{eqnarray}
\label{gama}
\gamma=\langle C_{x_1^{[1]}}(r)C_{x_1^{[2]}}(r)\rangle
\end{eqnarray}
where, $\langle\dots\rangle$ is time average. If one calculate $\gamma(\epsilon)$ as a function of $\epsilon$, then the systems are uncoupled if $\gamma=0$, but they are synchronized if $\gamma=1$ \cite{liu}.

Synchronization rate between two signals defined by kth variables in two coupled systems, $x_k^{[1]}(t)$ and $x_k^{[2]}(t)$ can be detected qualitatively by measuring a distance function parameter, $D_{x_k^{[1]},x_k^{[2]}}(t)=||x_k^{[1]}(t)-x_k^{[2]}(t)||$ \cite{pec,ram,ros,ros1}. The two systems are in (i) synchronous state if $D_{x_k^{[1]},x_k^{[2]}}(t)\rightarrow 0$, (ii) uncoupled state if $D_{x_k^{[1]},x_k^{[2]}}(t)$ fluctuates randomly, and (iii) transition state if the rate of fluctuation is about a constant value that is, $0~\langle~ D_{x_k^{[1]},x_k^{[2]}}(t)~\langle$ fluctuation (in uncoupled case).

The rate of synchronization can also be detected qualitatively by two dimensional recurrence plot of the corresponding variables in the two coupled systems\cite{pec}. The two systems are uncoupled if the points in plot are distributed randomly. However, if the two systems start coupled each other then the points in the plot start concentrating along the diagonal. The rate of synchronization is indicated by the rate of broadening of the points along the diagonal. If the two systems are strongly synchronized the points are just aligned along the diagonal, however, if the two systems are weakly synchronized, the points are scattered away a little showing a broaden diagonal line.
\begin{figure}
\label{}
\begin{center}
\includegraphics[height=230 pt,width=7.5cm]{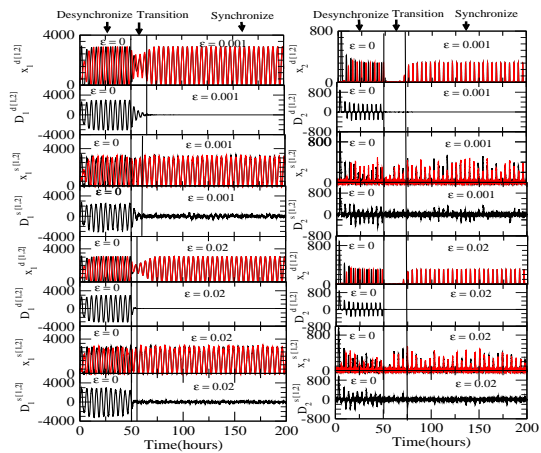}
\caption{The plots of p53 and Mdm2 dynamics of two cells diffusively coupled via NO at different $k_{NO}$ and coupling constant $\epsilon$ and their corresponding distance parameter $D^{[1,2]}$ dynamics showing different regimes, namely, desynchronized, transition and synchronized regimes. Coupling is switched on at 50hours.} 
\end{center}
\end{figure}

\section{Results}
We now first present the deterministic results by solving the set of differential equations (1)-(6) using standard 4th order Runge-Kutta algorithm for numerical integration \cite{press} as shown in Fig. 2 (upper two rows of the Fig. 2) in panels with superscripts $d$ on the variables. The parameter values taken for this single cell simulation are given in Table 2, and the value of $k_{NO}~(=k_9)$, creation rate constant, is allowed to vary. Since $NO\propto k_{NO}$, the value of $k_{NO}$ indicates the population of $NO$ in the system. This means that when the value of $k_{NO}$ is small the $NO$ present in the system is low and when the value of $k_{NO}$ increases, $NO$ present in the system is also increased. The results show that at lower value of $NO$ ($k_{NO}\le 0.005$), the two-dimensional plots of pairs of molecular species (proteins and their complexes) show fixed point oscillations indicating stabilization of the dynamics of these molecular species exhibiting normal behaviours of the respective molecular species in the system. However, further increase in $NO$ ($0.005\langle k_{NO}\le 0.1$) leads to the transition from fixed point oscillations to nearly limit cycle oscillation (limit cycle oscillation having certain thickness due to fluctuation in the dynamics) takes place. This indicates that $p53$ is activated with the increase in $NO$ showing the enability of $NO$ to cause DNA damage which leads to $p53$ activation \cite{hof}. If we further increase  $NO$ ($k_{NO}\rangle 0.1$), reverse transition i.e transition from the nearly limit cycle oscillations to fixed point oscillations takes place. This could be due to the fact that extremely increase in $NO$ can cause enormous decrease in Mdm2 and increase in p53 correspondingly in the system (i.e. too much toxic to the cell) leading to cell death \cite{xin}. So we have obtained two stabilization states in $p53$, one for normal like condition and the other for too much toxic leading to killing of cellular functions. In between these two stabilized states we get activated regime of $p53$ which consists of damped and sustained oscillatory behaviours depending on the values of $k_{NO}$. The term fixed point oscillation means oscillation death dynamics which is different from damped oscillation. Similar behaviour is obtained for dynamics of other molecular species.

We next present the stochastic results corresponding to the deterministic results by using the stochastic simulation algorithm (SSA) due to Gillespie \cite{gill} as shown in Fig. 2 (lower two rows) in panels with superscripts $s$ on the variables. The dynamics of each molecular species show noise induced and show similar behaviours as we have obtained in the deterministic case. The two stabilization and activation states are reached at faster rate (around 10$\%$ faster) in stochastic system as compared to deterministic case. In Fig. 3 we have found that for $k_{NO}$ = 0.001, the deterministic results show straight line (p53 is inactive) but the stochastic results show fluctuation (activated p53) due to noise.This shows that noise helps the system to reach various transition states significantly faster as compared to corresponding noise free system.
\begin{figure}
\label{}
\begin{center}
\includegraphics[height=150 pt,width=7.5cm]{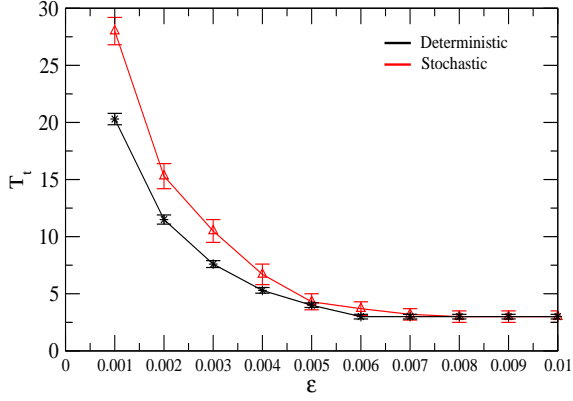}
\caption{Plot of transition time $T_t$ vs $\epsilon$ for both deterministic and stochastic systems with error bars.} 
\end{center}
\end{figure}

We then calculated permutation entropy for $p53$ time series data in deterministic system, $H_3^{x_1^{d}}$ based on the procedure described in the previous section and is shown in Fig. 3 in upper two panels: dynamics of $x_1^{[d]}$ is shown in uppermost panel and next panel shows corresponding $H_3^{x_1^{d}}$. Calculation of $H_3^{x_1^{d}}$ is done for embedded dimension $r=3$ with four distinct states ($S=4$) out of $r!$ permutations, size of the window $L=1000$ for different values of $k_{NO}$ ($[0.001-0.2]$). Since at low $k_{NO}~(0.001)$ $x_1^{d}$ shows fixed point oscillation (first stabilized state of $p53$) and the system is deterministic, the uncertainty in the system is minimized. Therefore, the corresponding $H_3^{x_1^{d}}$ to this $x_1^{d}$ dynamics shows minimized value (nearby zero) (Fig. 3 second uppermost panel). Then as $NO$  increases $(k_{NO}=[0.005-0.1])$, $x_1^{d}$ dynamics starts showing oscillatory behaviours (leading to activated state) with increasing amplitude but time period of oscillation approximately remain unchanged. This start of $p53$ oscillations leads to uncertainty in the dynamics that let $H_3^{x_1^{d}}$ increased which can be seen in the plot. If we increase the value of $k_{NO}$ further (corresponding to increase in $NO$), $H_3^{x_1^{d}}$ fluctuates with constant maximum level (remains the same for all $k_{NO}$ values) but with thicker points in $H^{x_1^{d}}$ dynamics. The thicker points in $H^{x_1^{d}}$ dynamics could be due to increase in uncertainty due to increase in activation. In the second stabilization state with excess $NO$, $H_3^{x_1^{d}}$ is constant at higher value as compared to the first stabilization state but with increase in fluctuation. Since increase $NO$ induce more $x_1^d$ via $x_2^d$ (increase stress in the system), it will induce more uncertainty in $x_1^{d}$ dynamics due to which stabilization occurs at higher uncertainty. Similar pattern is found in the case of $x_2^{d}$ ($Mdm2$) dynamics  as shown in 5th and 6th panels starting from uppermost in Fig. 3. 

We further calculated $H^{x_1^{s}}$ and $H^{x_2^{s}}$ for stochastic system for $x_1^s$ and $x_2^s$ respectively for various $k_{NO}$ and other parameters' values taken in the deterministic case and are shown in 3rd, 4th, 7th and 8th respectively in Fig. 3. The values of $H^{x_1^{s}}$ and $H^{x_2^{s}}$ are constant for a certain value of $k_{NO}$ with fluctuation about the constant value due to noise. As the value of $k_{NO}$ increases, activation of $x_1^s$ and $x_2^s$ increases however, the noise content in the dynamics helps to get stabilization quickly as compared to the deterministic case. This let $H^{x_1^{s}}$ and $H^{x_2^{s}}$ to decrease as $k_{NO}$ increases with increase in fluctuations due to increase in activation (increase in indeterminacy), and become stabilized with minimum $x_1^s$ and $x_2^s$ levels with minimum fluctuations. The dynamics of $x_5^d$ and $x_5^s$ (NO) in deterministic and stochastic systems with corresponding permutation entropies $H_3^{x_5^d}$ and $H_3^{x_5^s}$ are shown in 9th to12th panels in Fig. 3. 

We then study the behaviour of permutation entropy spectrum of stochastic time series by calculating it using three different permutation entropy calculations: first calculating it using Bandt and Pompe procedure \cite{ban} (indicated by black colour curve), second calculate M time series ensembles with different initial conditions, take average of these ensembles ($\{x_E:\frac{1}{M}\sum_{j=1}^{M}x_k^{j}, ~k=1,2,\dots, L_T\}$, where, $L_T$ is total length of the time series data), then we apply Bandt and Pompe procedure to calculate the permutation entropy $H_3^{E^s}$ of this $x_A^s$, and third we calculate permutation entropies ($H_3^1,H_3^2,...,H_3^M$) of $M$ time series data, then take average of these permutation entropy spectrums $H_3^{A^s}=\frac{1}{M}\sum_{j=1}^MH_3^j$. The averaging calculations reduce the fluctuations but the behaviour in stochastic system approximately remains the same as shown in Fig. 4. The behaviour of $H_3^{A^s}$ much better in agreement with stochastic single time permutation entropy as evident from the Fig. 4 and the value of permutation entropy decreases as $k_{NO}$ increases.
\begin{figure}
\label{}
\begin{center}
\includegraphics[height=300 pt,width=7.5cm]{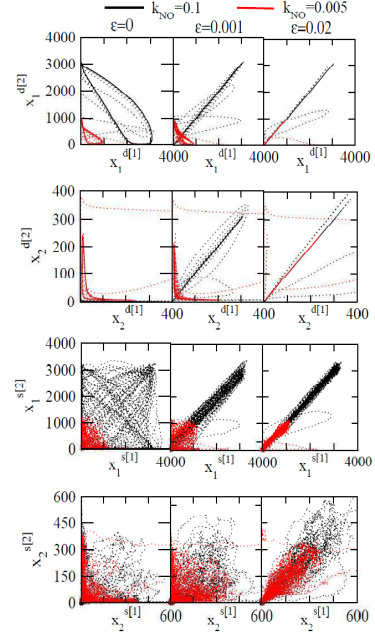}
\caption{2D-recurrence plots of p53 and Mdm2 of two cells diffusively coupled via NO at different $k_{NO}$ and coupling constant $\epsilon$. Both deterministic and stochastic results are presented.} 
\end{center}
\end{figure}

The deterministic steady state solutions of the single cell model can be obtained by taking $\frac{dx_i}{dt}=0$, $i=1,2,\dots,6$ of the deterministic equations (1)-(6) and solving for each variable from the steady state equations. We first solve for steady state solution of $x_{1D}$ variable by substituting and eleminating other variables using the equations to express $x_{1D}$ in terms of $x_{5D}$ which we found to be a quadratic equation in $x_{1D}$. Since the negative solution of this quadratic equation has no meaning, we take positive solution only. Since $x_{5D}^*=\frac{k_{NO}}{k_{12}}$ we have the solution for $x_{1D}^*$ given by,
\begin{eqnarray}
x_{1D}^{*}\sim\Gamma \sqrt{k_{NO}}\left(1+\frac{k_4k_{12}}{k_{NO}k_{10}}\right)^{1/2}
\end{eqnarray}
where, $\Gamma=\sqrt{\frac{k_3k_5k_{10}}{k_1k_2k_7k_{12}}(1+\frac{k_8}{k_6})}$ is a constant. It shows that as the increase in $k_{NO}$ the steady state of $p53$ is decreased. The first near normal steady state maintains at lower value of $p53$ which is hardly influenced by low value of $k_{NO}^*$ and the steady state of $p53$ is increased with increase in $k_{NO}$. Since $\frac{k_4k_{12}}{k_{NO}k_{10}}\langle 1$ (Table 2), it can be seen that $x_{1D}^{*}\propto\sqrt{k_{NO}}$.

Similarly, the steady state solution for $x_{2D}^*$ is obtained by solving the steady state equations, and is given by,
\begin{eqnarray}
x_{2D}^{*}\sim\frac{\Lambda}{\sqrt{k_{NO}}}\left(1+\frac{k_4k_{12}}{k_{NO}k_{10}}\right)^{-1/2}
\end{eqnarray}
where, $\Lambda=\sqrt{\frac{k_1k_2k_5k_{12}}{k_3k_7k_{10}}(1+\frac{k_8}{k_6})}$ is a constant. The $Mdm2$ steady state decreases as $k_{NO}$ increases which leads to the conclusion that first near normal stabilization of steady state of $p53$ maintains at larger value than the second stabilization of steady state of $p53$. Further, we also get that $x_{2D}^{*}\propto\frac{1}{\sqrt{k_{NO}}}$.
\begin{figure}
\label{}
\begin{center}
\includegraphics[height=150 pt,width=8cm]{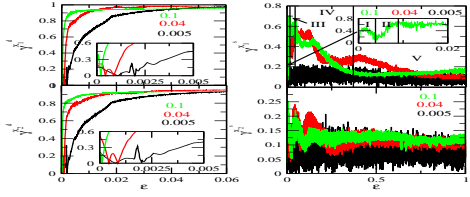}
\caption{Plots of the order parameter $\gamma$ as a function of $\epsilon$ for different values of $k_{NO}$. We present results of both deterministic and stochastic results.} 
\end{center}
\end{figure}

We then calculated steady state solutions of the variables $p53$ and $Mdm2$ for stochastic systems by applying the same procedure in the set of CLE (7)-(12) and solving for $x_{1CLE}^{*}$ and $x_{2CLE}^{*}$. Further we related deterministic and stochastic results for $p53$ which is given by,
\begin{eqnarray}
\label{cle1}
x_{1CLE}^{*}\sim x_{1D}^*\left[2-\frac{k_6}{2(k_6+k_8)}+\delta(V,\xi_1,\xi_2)\right]
\end{eqnarray}
where, $\delta(V,\xi_1,\xi_2)$ is the noise contribution to the deterministic result given by,
\begin{eqnarray}
\delta(V,\xi_1,\xi_2)&=&\left[1+\frac{k_6(k_5-2)}{2(k_6+k_8)}\right]\frac{\xi_1}{\sqrt{k_5V}}\nonumber\\
&&+\frac{\sqrt{k_5k_6k_8}}{\sqrt{2}(k_6+k_8)}\xi_2
\end{eqnarray}
The equation (\ref{cle1}) indicates that as $\delta$ (contribution of the noise in the stochastic systems) increases $x_{1CLE}^*$ also increases. The equation indicates that  $\frac{x_{1CLE}^{*}-x_{1D}^*}{x_{1D}^*}=1-\frac{k_6}{2(k_6+k_8)}+\delta \rangle 0$ and therefore $x_{1CLE}^{*} \rangle x_{1D}^*$. This means that noise in stochastic system help the system to raise molecular population by probably enhancing the molecular interaction in the systems.

Similarly, the steady state solution of $Mdm2$ in stochastic system is obtained by solving the steady state equations of CLEs. It is given by,
\begin{eqnarray}
x_{2CLE}^*\sim x_{2D}^*\left[2-\frac{k_6}{2(k_6+k_8)}+\delta(V,\xi_1,\xi_2)\right]
\end{eqnarray} 
Similar result is obtained as in the case of $p53$ and noise helps in getting stabilizations and activation early as compared to deterministic case.
\begin{figure}
\label{}
\begin{center}
\includegraphics[height=220 pt,width=8cm]{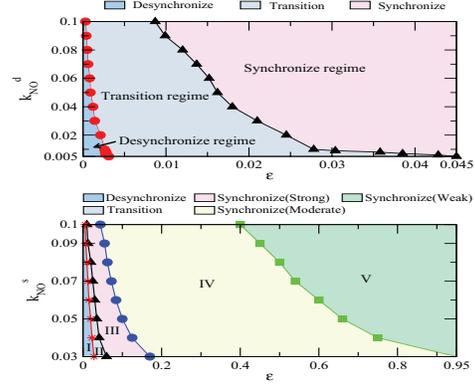}
\caption{Phase diagram of deterministic system in the parameter space ($\epsilon-k_{NO}$) indicating desynchronized, transition and synchronized regimes respectively.} 
\end{center}
\end{figure}

\subsection{Results of coupled stress cells}

We now consider two such identical systems diffusively coupled with $NO$ as coupling molecule. This molecular coupling can be done by constructing a larger system defined by $\vec{x}(t)=(x_1^{[1]},x_2^{[1]},\dots,x_6^{[1]},x_1^{[2]},x_2^{[2]},\\ \dots,x_6^{[2]})^{-1}$ where the identical systems are its subsystems and then introducing two coupling reactions, $x_5^{[1]}\stackrel{\epsilon}\longrightarrow x_5^{[2]}$ and $x_5^{[2]}\stackrel{\epsilon^\prime}\longrightarrow x_5^{[1]}$. The system can be described by a set of 12 coupled differential equations with extra coupling terms $\epsilon (x_5^{[2]}-x_5^{[1]})$ and $\epsilon (x_5^{[1]}-x_5^{[2]})$ added to the differential equations with $x_5^{[1]}$ and $x_5^{[2]}$ respectively, where $\epsilon=\epsilon^\prime$ is taken. Putting all the rate constant values in the coupled sub-systems to be the same, we solve the differential equations of the deterministic system numerically for various values of coupling constant, $\epsilon$. The results for $p53$ and $Mdm2$ for the coupled systems are shown in Fig. 5 as time series of the coupled systems and their corresponding $D_{i}^{[1,2]}~(i=1,2)$, where superscript with d is for deterministic and superscript with s is for stochastic and the coupling is switched on at 50 hours with different values of $\epsilon$ ([0.001-0.02]). The results show that there are three distinct states, namely, desynchronized (the two systems are uncoupled and therefore $D_{i}^{[1,2]}$ fluctuates randomly), transition (time to reach synchronized state from desynchronized state and $D_{i}^{[1,2]}$ weakly fluctuates) and synchronized states ($D_{i}^{[1,2]}$ become constant with small fluctuation about it). It is also seen that transition time decreases as $\epsilon$ increases which is evident both from time series data as well as  from $D_{i}^{[1,2]}$ in Fig. 5. 

 Again we study the two coupled systems with various  values of $\epsilon$ and calculated the approximate transition time $T_t$. $T_t$ is the time taken to reach from transition to synchronized state after coupling is switched on. In the deterministic case we could get the synchronization faster as compared to the stochastic system. Synchronization is achieved if the dynamics of the corresponding variables  are same giving  $D_{i}^{[1,2]}$ $\longrightarrow$ 0, which is easily seen in Fig 5 panel 2nd and 6th.The behaviour of $T_t$ as a function of $\epsilon$ is shown in Fig 6. Here error bars for each $\epsilon$ values are calculated by averaging 10 different initial values for all variables. The Fig 6 shows that $T_t$ is an exponentially decaying function of $\epsilon$  which is given by  $T_t(\epsilon$)=$Ae^{-\alpha\epsilon}$ + B , where A and B are constants. Here we can see that $T_t$ decreases as $\epsilon$ increases up to some critical value that is $\epsilon$ = 0.007, after which $T_t$ remains constant.

We then switch on the coupling at 0 hour and the deterministic results of the coupled systems are shown as recurrence plots in the planes ($x_1^{d[1]},x_1^{d[2]}$) and ($x_2^{d[1]},x_2^{d[2]}$) respectively for three different values of $\epsilon=0,0.001,0.02$ and $k_{NO}=0.1,0.005$ respectively in the first two upper sets of panels in Fig. 7. The two oscillators are found to be uncoupled for $\epsilon=0$ both for small and large concentration levels of $NO$. The rate of synchronization starts increasing as the value of $\epsilon$ increases indicated by the rate of concentration of the points towards the diagonals of the plane. The variables $x_1^{d[1]}$ and $x_1^{d[2]}$ of the two systems become strongly synchronized when $\epsilon\ge 0.02$ both for $k_{NO}$ values. However the rate of synchrony is slow for lower $k_{NO}$ value as compared to that of higher values of $k_{NO}$ as evident from the plots. 
\begin{figure*}
\label{}
\begin{center}
\includegraphics[height=250 pt,width=13cm]{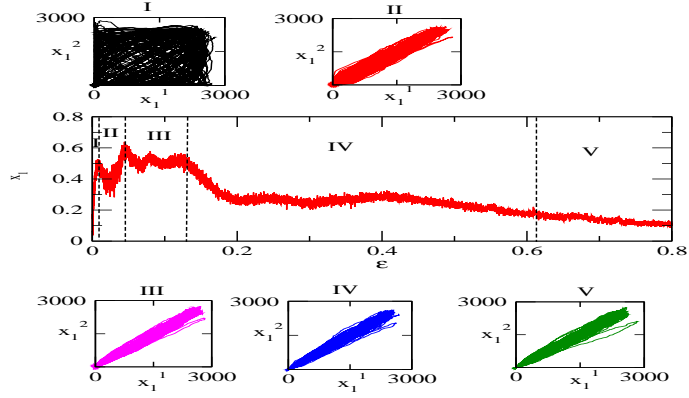}
\caption{Plots of $\gamma$ as a function of $\epsilon$ in stochastic system indicating different regimes of synchronization.} 
\end{center}
\end{figure*}

The same pattern of synchronization in $p53$ and $Mdm2$ which we have found in deterministic case is obtained in the stochastic case also for the same $\epsilon$ and $k_{NO}$ values. However, the rate of synchronization is much stronger in deterministic case as compared to stochastic case as the spreading of the points from diagonal in the respective planes in deterministic case is much thinner than same spreading of the points in stochastic case. This means that the role of the noise in coupled systems is to destruct the synchronization.

We then present the stochastic results corresponding to the deterministic results with superscript 's' in Fig. 8 (right two panels) and Fig. 9 (lower panel). We get similar results as in the deterministic case except that $\gamma^{x_1^{s}}$ and $\gamma^{x_2^{s}}$ versus $\epsilon$ have larger fluctuations induced by noise and synchronization occur at larger $\epsilon$ values as compared to deterministic case as shown in right panels of Fig. 8. We then extended the range of $\epsilon$ to see the effect of excess $NO$ diffusion in the coupled system. We now found five regimes in the phase diagram plotted in ($\epsilon-k_{NO}$) plane shown in lower panel of Fig. 9 and Fig. 10. The regimes $I$, $II$ and $III$ are desynchronized, transition and strongly synchronized regimes respectively. The regimes $IV$ and $V$ are the regimes where excess $NO$ present in the stochastic system induce decrease in synchronization rate. When the excess of $NO$ is moderate, the rate of synchronization is found to be reduced by $30\%$ from the strongest synchronization value for a small range of $\epsilon$ as shown in regime $IV$ in the Fig. 9 and Fig. 10. It shows the toxic nature of $NO$ and there is a competition between synchronization and toxic activities of $NO$ in the coupled system. Further, if excess of $NO$ level is stronger, then $\gamma^{x_1^{s}}$ is reduced drastically again and become almost constant. This leads us to claim that if the excess of $NO$ is very strong, the toxic activity of $NO$ could dominate over its synchronization activity and may lead to cell death. Similar pattern is obtained in the case of $Mdm2$. Further it is to be noted that as the value of $k_{NO}$ increases, the shifting of the system from desynchronized ($I$) to transition state ($II$) and then to strong synchronization state ($III$) is achieved at smaller value of $\epsilon$. However in the excess of $NO$ regime ($IV$ and $V$) the toxic activity dominates the synchronization activity leading to decrease in the rate of synchronization as indicated by regime ($IV$ and $V$) of Fig. 10.

Next we investigate the amplitude variation $A^{x_1}$ of $x_1$ as a function of $k_{NO}$ to understand the impact of $NO$ and its activity on $p53-Mdm2$ regulatory process as shown in Fig. 11 both for deterministic and stochastic systems. We found three distinct types of oscillatory behaviours and their transitions in the ($A^{x_1}-k_{NO}$) phase diagram. Initially as we increase the value of $k_{NO}$, $A^{x_1}$ starts increasing which falls in damped oscillation regime (could be increase in $p53$ activation due to increase in $NO$). If we increase $k_{NO}$ further, transition from damped oscillation to sustain oscillation regime takes place that could be due to $p53$ activation by $NO$ is strongest. Now if we again increase $k_{NO}$ value, then $A^{x_1}$ starts decreasing which is given by transition from sustain to damped oscillation again. This shows the increase in toxicity in the cell due to excess of $NO$. If we increase $k_{NO}$ further, $A^{x_1}$ is drastically decreased and become constant leading to transition from damped oscillation to fixed point oscillation behaviour. This could be due to excess $NO$ induced more toxic leading to cell death.
\begin{figure}
\label{}
\begin{center}
\includegraphics[height=200 pt,width=7.5cm]{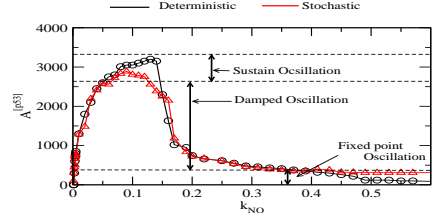}
\caption{Variation of amplitude of p53 as a function of $k_{NO}$.} 
\end{center}
\end{figure}

\subsection{Stability analysis of coupled system}

Now we calculate steady state solution of the coupled system to study the impact of $NO$ concentration level on the activation and stabilization of $p53$. This can be done same as we have done in the case of single system. The solution for $NO$ in the coupled system, $x_5^{*[1]}$ and $x_5^{*[2]}$ is given by,
\begin{eqnarray}
x_5^{*[1]}=x_5^{*[2]}exp\left[-{\frac{k_d-k_{NO}}{\epsilon}}\right]
\end{eqnarray}
This equation indicates that $x_5^{*[1]}=x_5^{*[2]}$ when $k_d=k_{NO}$. We again solve for $x_1^{*[1]}$ and $x_1^{*[2]}$ from stability equations and is given by,
\begin{eqnarray}
\label{del}
\frac{\Delta x_1^*(\epsilon,k_{NO})}{x_1^{*[2]}}=\frac{k_{NO}}{2\epsilon x_5^{*[1]}}\left(\frac{1-\frac{k_{11}}{k_{NO}}}{1+\frac{k_4}{k_9x_5^{*[1]}}}\right)
\end{eqnarray}
where, $\frac{\Delta x_1^*(\epsilon,k_{NO})}{x_1^{*[2]}}=(x_1^{*[1]}-x_1^{*[2]})/x_1^{*[2]}$ is the rate of $NO$ concentration diffused from first subsystem to the second subsystem. The two subsystems in the coupled system will have same stability state (stationary fixed point) when $\frac{\Delta x_1^{*[1]}}{x_1^{*[2]}}=0$. The equation (\ref{del}) further indicates that $\frac{\Delta x_1^{*[1]}}{x_1^{*[2]}}\propto\epsilon^{-1}$ and $\frac{\Delta x_1^{*[1]}}{x_1^{*[2]}}\propto\frac{1}{x_5^{*[2]}}$ such that the the coupled system will get stabilized (the two subsystems reaching at same stability state) when $\epsilon\rightarrow\infty$ and $x_5^{*[1]}\rightarrow\infty$.

In the same way we calculated the steady state solution by relating $x_2^{*[1]}$ with $x_2^{*[2]}$, and is given by,
\begin{eqnarray}
\label{del1}
\frac{\Delta x_2^*(\epsilon,k_{NO})}{x_2^{*[1]}}=\frac{k_{NO}}{2\epsilon x_5^{*[1]}}\left(1-\frac{k_{11}}{k_{NO}}\right)
\end{eqnarray}
where, $\frac{\Delta x_2^*(\epsilon,k_{NO})}{x_2^{*[1]}}=(x_2^{*[2]}-x_2^{*[1]})/x_2^{*[1]}$ is the rate of $x_2^{*[2]}$ concentration diffused from second subsystem to the first subsystem. The two subsystems reach at the same stability state when $\epsilon\rightarrow\infty$ and $k_{NO}\rightarrow\infty$.

\section{Discussion}
The possible impact of nitric oxide molecule on the $p53-Mdm2$ regulatory network in single cell as well as in coupled cells are investigated on a model developed based on various experimental reports. Nitric oxide molecule is being created due to protein-protein interaction inside the cell and is believed to be toxic in normal cells. Hence the nitric oxide is maintained at low, which is controlled by various sub-cellular networks. However, various stress conditions induced in the cell enhance the creation of nitric oxide and directly influence ($p53-Mdm2$) network via $Mdm2$. This leads to the activation of $p53$ molecule in the network which we get the evidence in our work. In the single cell model, the $p53$ behaves as nearly normal keeping it low and stabilized when $NO$ is low. If the $NO$ is increased significantly, $p53$ protein is activated indicated by its oscillatory behaviour. However excess $NO$ prohibits the $p53$ activation due to too much toxic induced and $p53$ stabilized but at higher value as compared to normal cell. This may lead to cell apoptosis. 

Another important property of $NO$ is that it has been known as one of the most capable signaling molecules which can diffuse across the cell membrane. We consider this signaling molecule as synchronizing agent which can diffusively couple any two cells and study various behaviours in the coupled cells. The cells behave as uncoupled or non-interacting individuals if the coupling strength is small. However if the coupling strength is stronger then the cells start interacting each other by passing information via synchronizing molecule but may not strong enough to get complete synchronization (transition regime). If the coupling strength is strong enough then the two cells get synchronized. It is also seen that synchronization is reached much faster in deterministic case than in stochastic system giving the destructive role of noise in achieving synchronization. Since increase in coupling strength of $NO$ means increase in $NO$ diffusion in and out of the cell which in turn induce more toxic in the cell. This excess of $NO$ diffusion leads to decrease in rate of synchronization even if coupling strength is increased.

There are various issues to be solved in future for example information transmission and receiving among a large number of cells and spatio-temporal dependence of synchronization. Since the $p53$ protein is hugely connected hub, various influences of signaling molecules from various sub-networks need to be considered simultaneously.

\section*{Acknowledgments}
This work is financially supported by University Grant Commission (UGC), India and carried out in Centre for Interdisciplinary Research in Basic Sciences, Jamia Millia Islamia,New Delhi,India.

\end{document}